# “X-problem of value three”, definition

A. Kornyushkin
kornju.hop.ru, kornju@mail.ru

**Abstract**
The “X-problem of number 3” for one dimension and related observations are discussed.

## 1 Introduction

An extensive area of "Automata Machine Configuration 2x3" precedes the Numbers theory. There is one interesting problem related to Automata Machine Configuration 2x3. It is all about one surprising observation that shows the relationship between the N- dimensional space and the properties of number 3. Below, we formulate definitions for the problem in one dimensional case.

## 2 Reversible cellular machine on the graph

**Let us prove the three not complex theorems and give one impotant definition.**

Consider mixed, finite graph $G=(V,E,A)$. We are using the definition from the following resource: http://en.wikipedia.org/wiki/Graph_%28mathematics%29. Below, we will use only this definition and no other developments from the graph theory.

Here $V=\{v_i\}$ is the node of graph, $E=\{e_i\}$ are directed edges, and $A=\{a_i\}$ are undirected edges. Our graph is ordinary and, therefore, does not contain loops and multiple edges. Briefly, a directed edge is depicted with the arrow from one node to another, while undirected edge is depicted with a simple junction between the two nodes. The junction can be otherwise depicted using two arrows: one arrow leads from node 1 to node 2 and another leads from node 2 to node 1. Thus, nodes 1 and 2 in our combined graph can not bind undirected and directed edge together. This is pointless. directed communication is already there. (See Fig. 1).

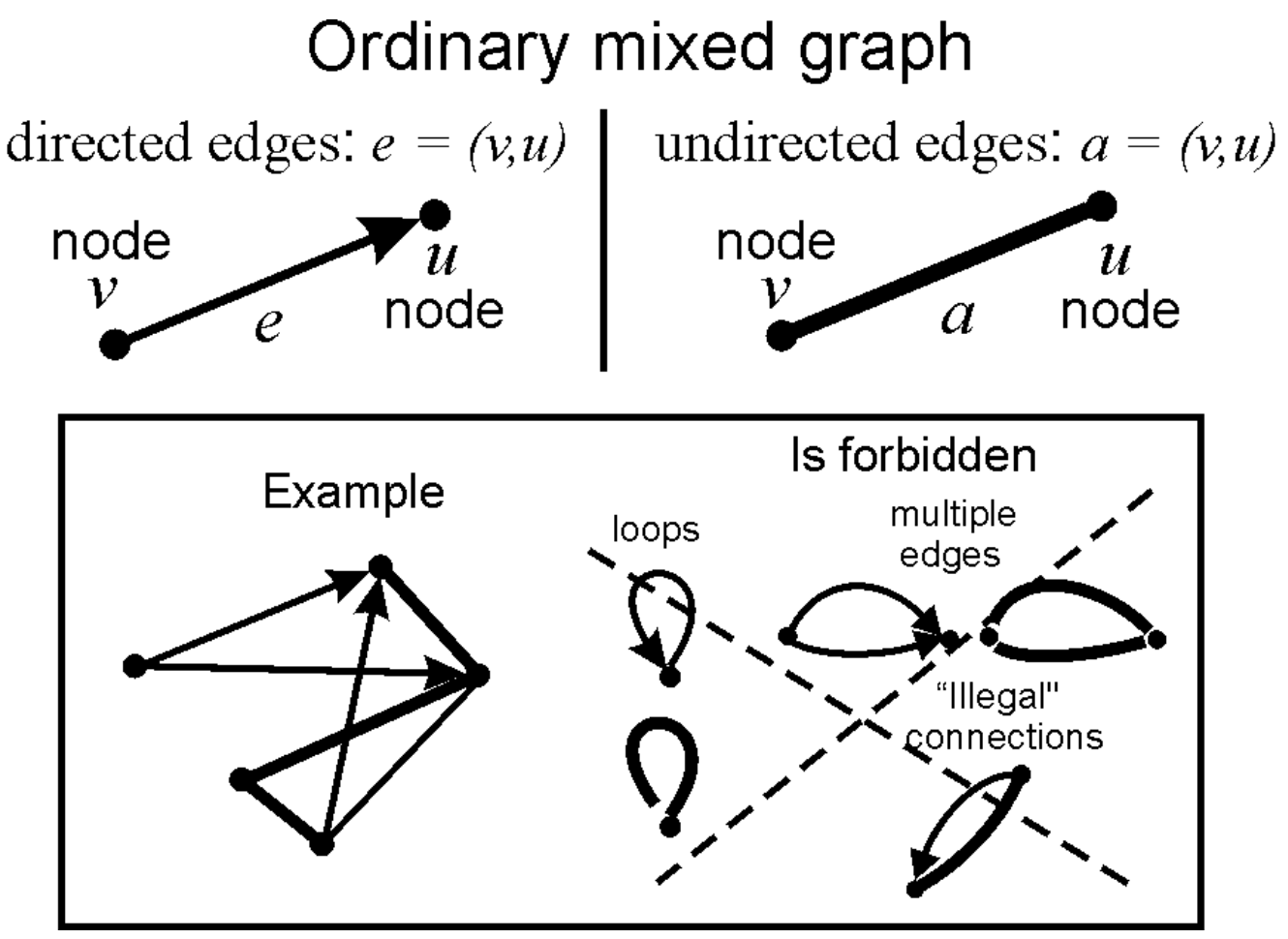

Fig.1.

**Definition 1.**

The graph is called Super Weak computable when there is a closed route through all the nodes using both the directed and undirected edges

$$v_0 \overset{a_0 \vee e_0}{\longrightarrow} v_1 \overset{a_1 \vee e_1}{\longrightarrow} \ldots \overset{a_k \vee e_k}{\longrightarrow} v_0$$

Tnis definition can be rephrased as follows: using either directed or undirected edges or a combination of directed and undirected edges one can pass through all the nodes of the graph and come back to the starting node.

**Definition 2.**

The graph is called Weak computable when there is a route

$$v_0 \overset{a_0}{\longrightarrow} v_1 \overset{a_1}{\longrightarrow} \ldots \overset{a_k}{\longrightarrow} v_k$$

following which one can also pass through all the nodes using *only undirected* edges. So, it is clear: when the graph is Weak computable, it is also Super Weak computable. (See Fig. 2).

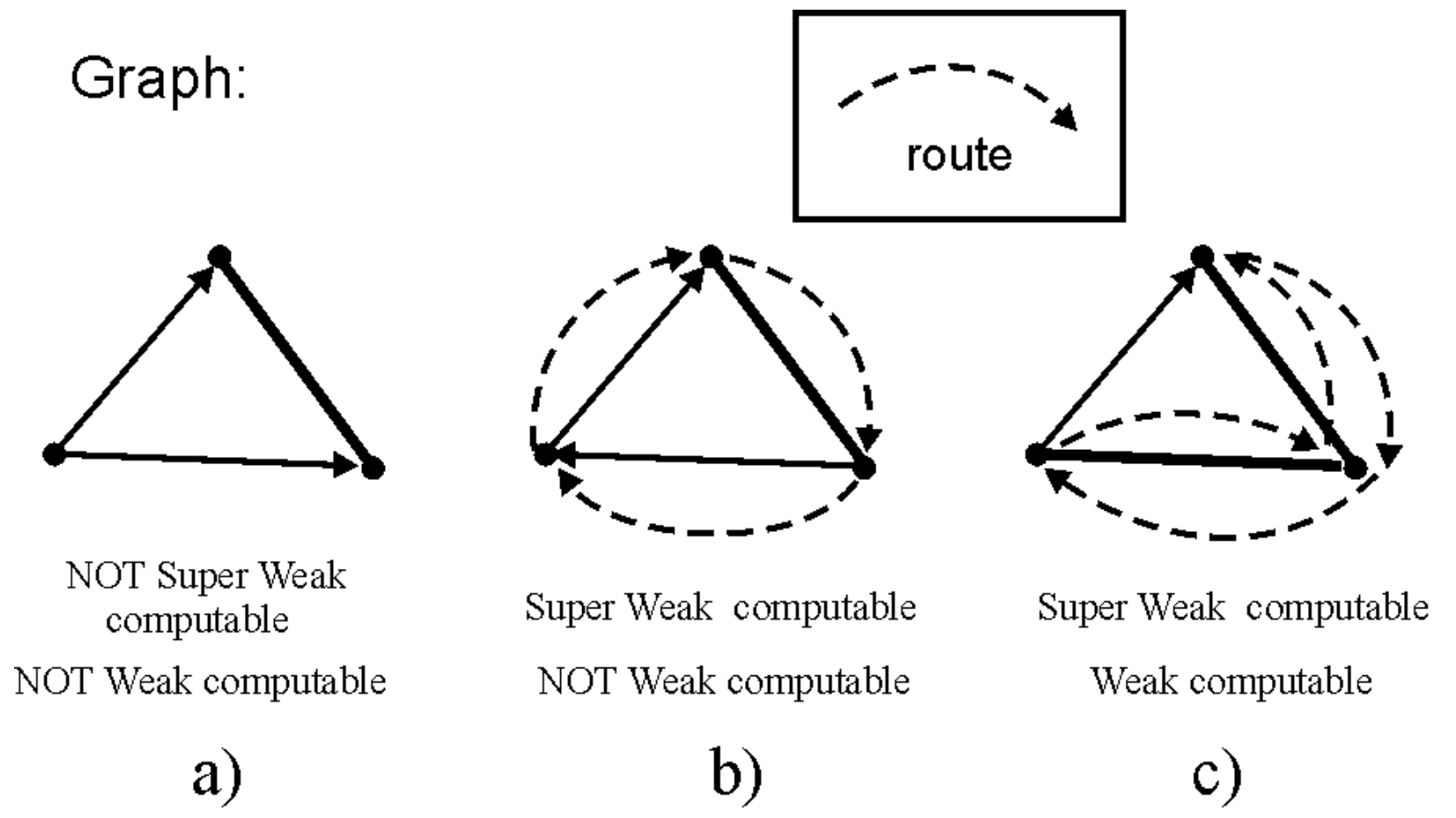

Fig. 2.

Let the nodes of the graph be of three colors (A, B, and C).

$$v_i \in \{A, B, C\}$$

Further, we introduce two definitions

**Definition one: Transliteration** of graph $G$ $(\tilde{G})$ is defined as the change in a color map of graph G where all the nodes of graph G colored in C are recolored into color B, and all nodes colored in B are recolored into color C.

$$G(B \Leftrightarrow C) \equiv \tilde{G}$$

**Definition two: Complement** of graph $G$ $(\overline{G})$ is defined as the change in a color map of graph G where all nodes colored in A are recolored into color B, and all nodes colored in B are recolored into A. Usually, this **complement** transformation rule is applied to the cases where there are no nodes colored in C.

$$G(A \Leftrightarrow B) \equiv \overline{G}$$

Let us enter the rule of transformation of colors in graph G:

$$G_t \to G_{t+1}$$

This rule depends on the existence of *at least one node colored in C* in the end of directed or undirected edges all of which start from the same node $v_i$. (We call this condition as P-condition).

So, if P=0 (i.e., there are no nodes colored in C) then one can make following recoloring

$$P = 0 \ (I) \Rightarrow \begin{array}{l} v_t = A \Rightarrow v_{t+1} = A \\ v_t = B \Rightarrow v_{t+1} = C \\ v_t = C \Rightarrow v_{t+1} = B \end{array}$$

Further we refer to this recoloring as rule I.

If P=1 (opposite situation), we realize following recoloring rule

$$P = 1 \ (II) \Rightarrow \begin{array}{l} v_t = A \Rightarrow v_{t+1} = C \\ v_t = B \Rightarrow v_{t+1} = A \\ v_t = C \Rightarrow v_{t+1} = B \end{array}$$

Further we refer to this recoloring as rule II. We note here that this recoloring is some kind of reversible cellular automation from [1].

**Theorem 1** (about reversability**).**

$$\tilde{G}_t \to \tilde{G}_{t-1}$$

**Proof.**

To prove the above theorem, it is enough to consider all five events (shown by numbers in circles) from Fig. 3.

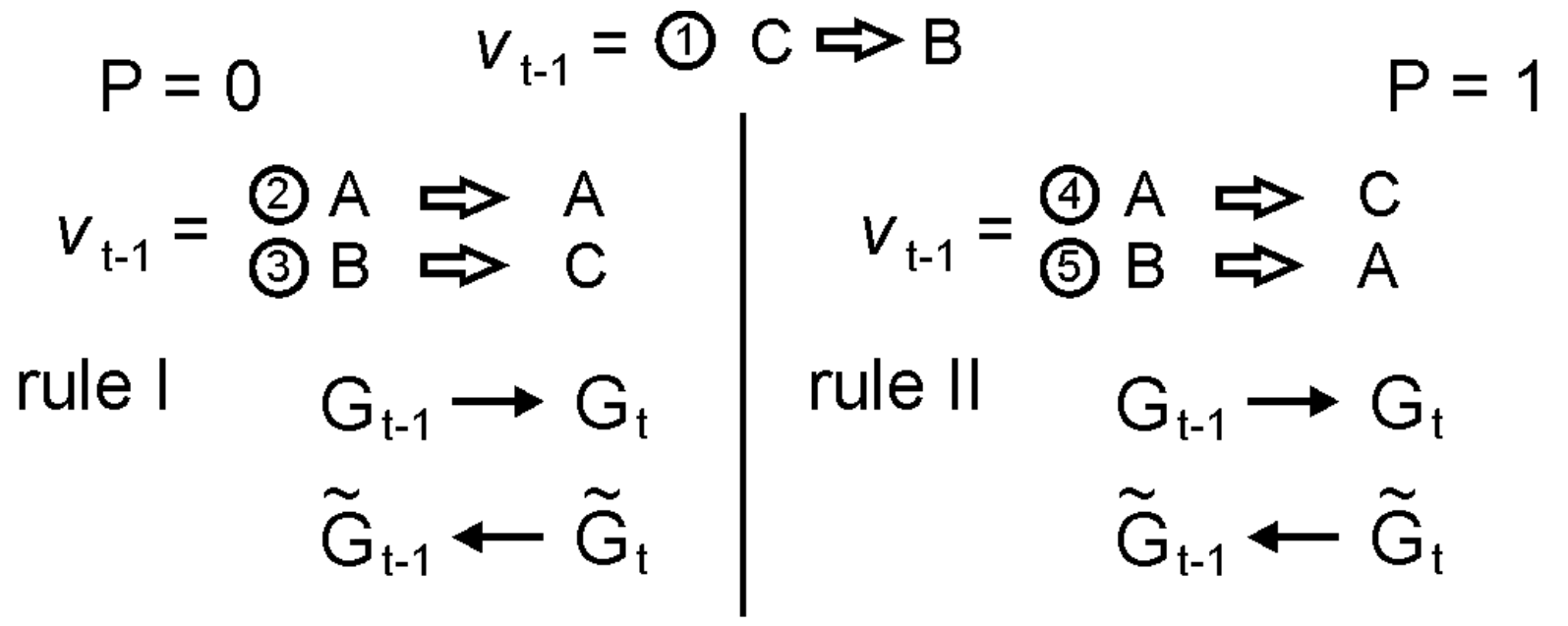

1) Node of color C always transfer into node of color B

2) Node of color B can be formed only by node of color C

Fig. 3.

Consider case (1). This case is obvious. If we do transliteration, then CB will be BC.

Consider case (2). Recoloring rule is rule I, so there are no nodes colored in C among the nodes of the graph where directed (or undirected) edges leads from node $v_{t-1}$. Therefore, in the derivative of this graph (i.e., in the graph that emerges after recoloring) there are no nodes colored in B. (Note here that the nodes colored in B can be formed only from the nodes colored in C). Consequently, in the transliterated graph, it will be no C nodes among all its nodes. This means that the transfer rule is still rule I. Also node $v_t$ in graph $\tilde{G}_t$ will be colored in A in accordance with the rule I.

The rest three cases can be considered in the same way. The theorem is a simple consequence of two statements shown in Fig. 3.

Assume that recoloring of graphs starts at time t = 0. Also assume that the nodes have two colors: A and B at the start. This means that during the next step (there are no nodes colored in C; so the rule I is applied to the entire graph) the graph will be converted into transliteration of itself, consisting only of A and C colors. (All nodes colored in B will transformed into nodes colored in C, whereas all nodes colored in A will stay the same) . So, the graph will go along reversed-time path (see Theorem 1). In the steps taking place at time $t = n^* = -n + 1$, the graph will be the transliteration of the same graph in time $n$ (see Fig. 4).

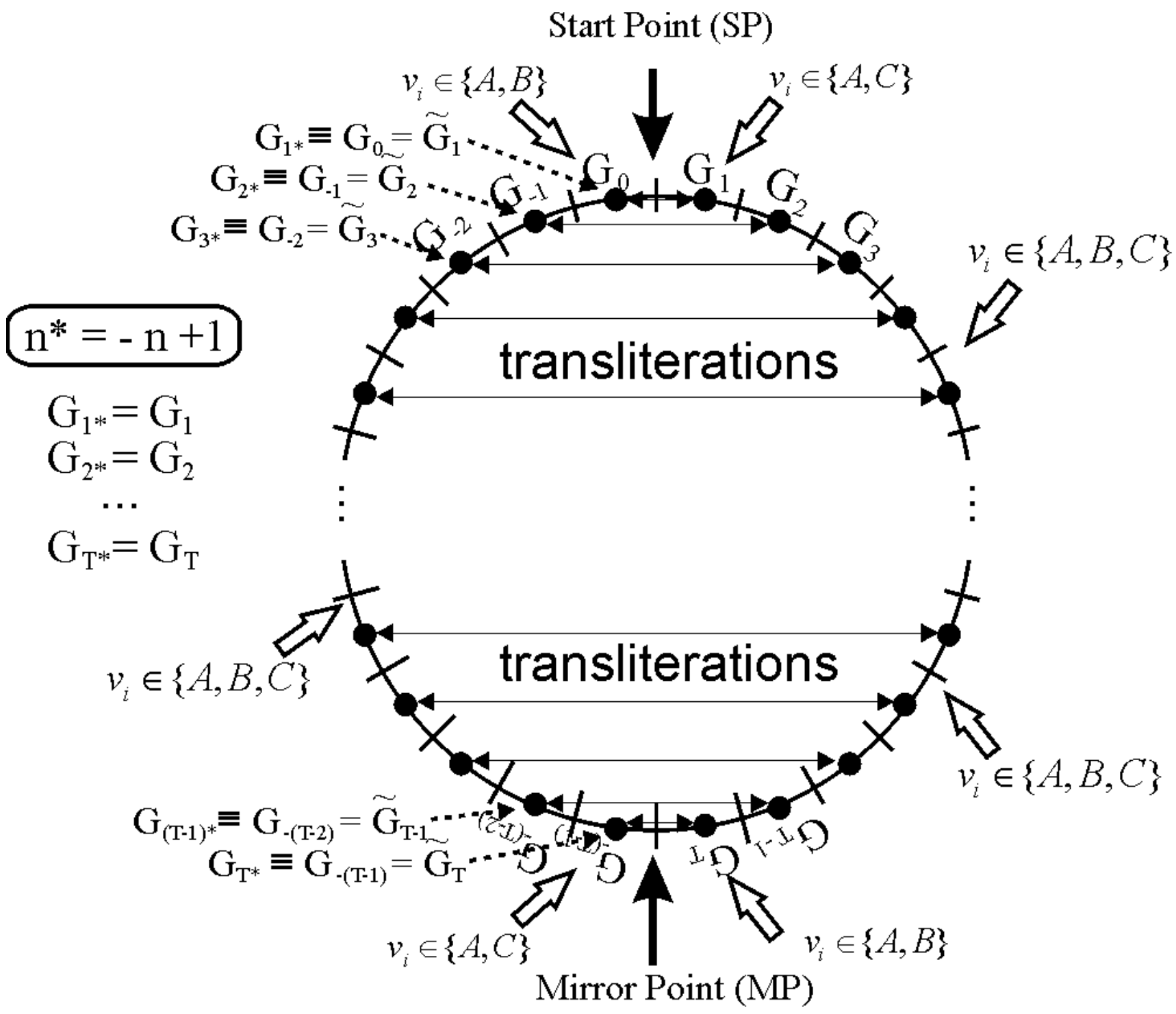

Fig. 4. Motion of recolouring of graph G.

Consider the time point located between the states of graph at time 1* and time 1 as the Start Point (SP) and continue the motion in two directions: let us called it "forward" time and "reversed" time. Due to the fact that our graph defined as finite, these two motions will meet at time T* - T somewhere at point $G_{T^*}$, where $G_{T^*}$ is the transliteration of $G_T$. Lets us call this point as Mirror Point (MP), and call T as **period** (or period of return). Here we use the main property of reversible finite Automats stating that any finite reversible Automat "is obligated to pass through its starting point". That is quite natural. If any recolored graph has only one predecessor (and it is easy to find this predecessor: let us make transliteration, then one step forward, then make transliteration again), then any recoloring of graph is cyclic. Also there is no any "branch" that merges with this "cycle".

If T > 2, then the Mirror Point exists. In this case, full period of return (2T) is an even number. Otherwise, the graph in the Mirror Point is transliteration of itself. This means that all nodes of graph are colored in A, and graph simply stays the same.

Let us prove following theorems.

**Theorem 2.** (About coloring of the Mirror Point).

Let T > 2 and the graph is Super Weak computabe. Let the Start Point be the state between the graph that has colors A and B ($G_{1^*}$) and the graph that has colors A and C ($G_1$). Then the Mirrow Point is the state between the graph that has colors A and B ($G_T$) and the graph that has colors A and C ($G_{T^*}$).

**Proof.**

If at moment $T$ of time graph $G_T$ has at least one node colored in C, then the original node where the edge leading to the node colored in C starts must be also colored in C. Recoloring rule II can not be in MP. It means that nodes colored in A are transformed into nodes colored in C, and the nodes colored in B are transformed into nodes colored in A. Yet, none of the nodes is transliteration of each other. And moving through all the nodes along the route leading to "another side", we can make a conclusion that the all graph nodes are colored in C. The graph state changes from "all nodes are C " to "all nodes are B ", and from "all nodes are B" to all nodes are C"(T< 3). Here, we immediately see the contradiction. So theorem 2 is proven.

From this theorem, follows the theorem of *extisting of number $\lambda$.*

**Theorem 3.**

If our graph is Weak computable, any $v_i$ of its nodes when moving from SP to MP

$$v_{i,t=1} \rightarrow v_{i,t=2} \rightarrow \ldots \rightarrow v_{i,t=T}$$

is transformed into equal number of nodes colored in A ($N_A$), B ($N_B$), and C ($N_C$), moreover, the number of nodes colored in B is equal to the number of nodes colored in C.

$$N_B = N_C \equiv N_{BC}$$

Let us introduce parameter $\lambda = N_A - N_{BC}$.

**Proof.**

Let us start from arbitrary node, move around all the nodes in the graph following undirected edges, and then return to the initial node. Every new step means "plus one" in our new numeration. It does NOT MATTER that some nodes will be passed tow or more times (see Fig. 5).

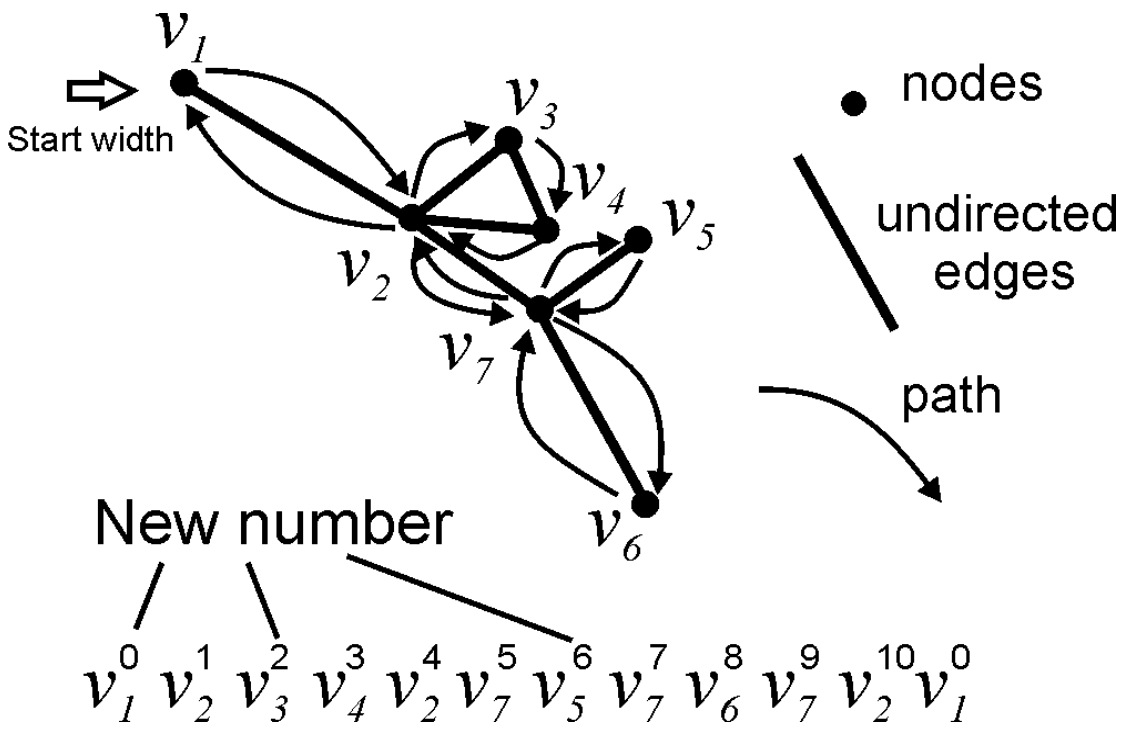

Fig. 5. Our "numeration" of the nodes.

Let us organize the node along a single horizontal line (circle). The most important thing in our numeration, that to the left and right sides of any node there will be a node that has a "tie" with our node through an undirected edge. Let us look at Fig. 6,

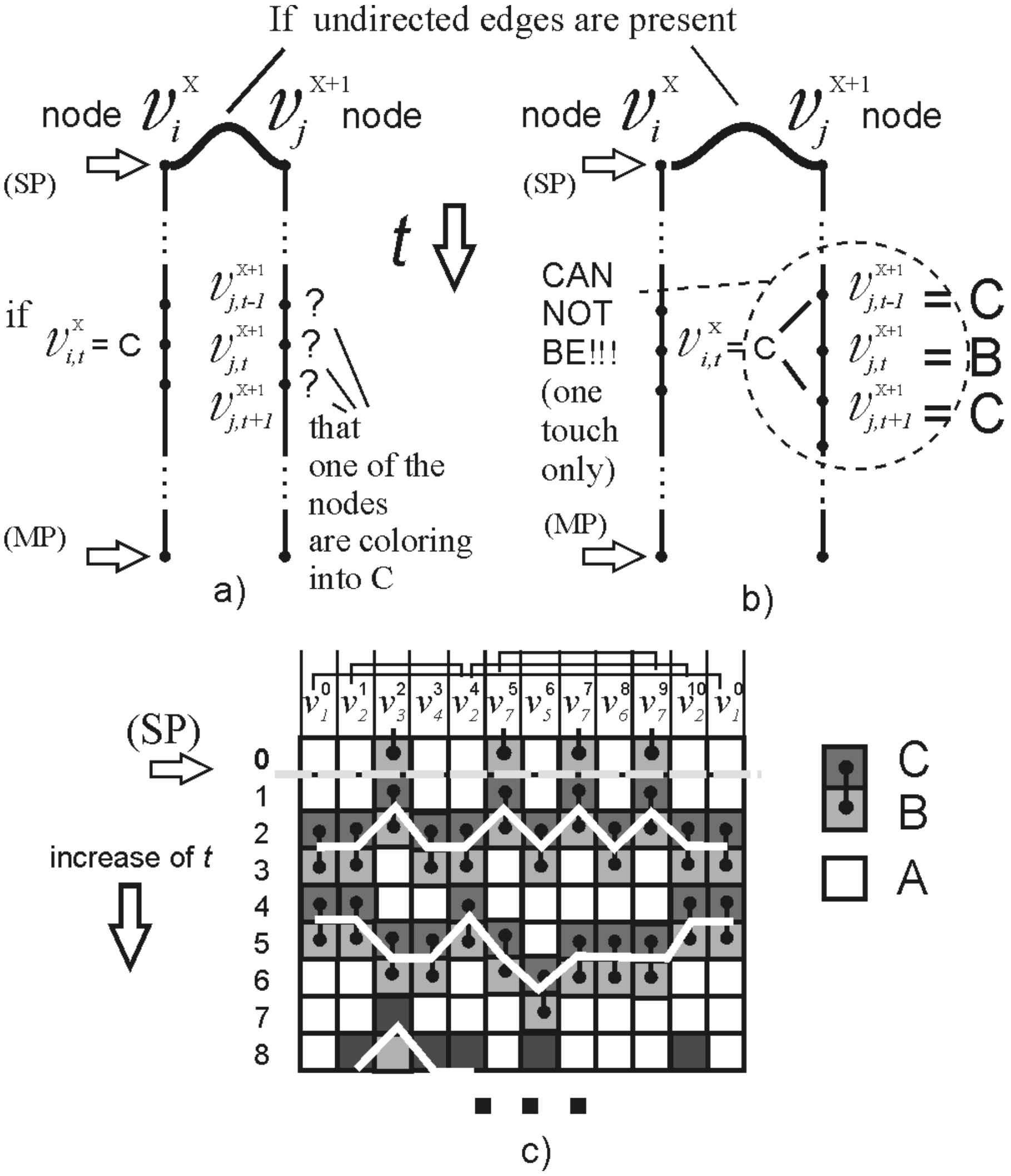

Fig. 6. Illustration of Theorem 3.

Let us look at the Fig. 6a. Due to existence of the "directed half" edges leading from node $v^{x+1}_t$ to node $v^x_t$ and due to the fact that $v^x_t$ is colored in C, one of the nodes $v^{x+1}_{t-1}$, $v^{x+1}_t$, $v^{x+1}_{t+1}$, will be also colored in C. Let us pretend that it is not the case and ask which color will be for the node marked by a central question mark (node $v^{x+1}_t$) in Figure 6a? If this node is colored in A, then the node beneath it (node $v^{x+1}_{t+1}$) is colored in C. (Note that transfer rule here is rule II). This node cannot be colored in color B, either. Then node $v^{x+1}_{t-1}$ above it is colored in C. But node $v^{x+1}_t$ cannot be colored in C. So, we have a contradiction here.

And we also note that a "double touch" is also not possible! $v^{x+1}_t$ = B cannot be transformed into $v^{x+1}_{t+1}$= C. Here a transfer rule must be rule II. In this case, $v^{x+1}_{t+1}$ must be colored in A. So, again we see a contradiction.

**That is, double-touch on either the right side or the left (our edge is undirected) is impossible!** So, when moving from SP to MP, our figure consists of non-intersecting CB bands (see Fig. 6c).

Thus, the theorem is proven.

Take Weak computable graph. Consider graph $G_{1^*}$, consisting of colors A and B and having its own $T$ and $\lambda$ and complementary graph $\overline{G}_{1^*}$: graph $\overline{G}_{1^*}$ comprises only colors A and B and has its own $\overline{T}$ and $\overline{\lambda}$. We assume that $\overline{T} \geq T$ , otherwise, just rename $G_{1^*}$ and $\overline{G}_{1^*}$.

Let us take arbitrary node $v$ for first graph $G_{1^*}$ and consider the entire array of colors $v(t)$ the node passes on its way from SP to MP. The size of this array is $T$.

Let us enter four arrays $a_v(t)$, $b_v(t)$, $c_v(t)$, $\lambda_v(t)$; where t = 0, 1, 2...T–1.

Now, $a_v(t) = 1$, if $v(t)$=A, and $a_v(t)=0$, otherwise. The same rules are true for $b_v(t)$, $c_v(t)$.

So

$$\lambda_v(t) = \sum_{0 \leq p \leq t}^{p} (2a_v(p) + b_v(p) + c_v(p) - 1)$$

Let us define three arrays $C_v(k)$, $A_v(k)$, $f_v(k)$. (array $f$ is called “integral phase”).

$C_v(k)=0$; but if we can find $t$ which makes value $k=\lambda_v(t)/2$ an integer number and therefore $v(t)=C$, then $C_v(k)=1$.

If $C_v(k)=1$, then we can write that $f_v(k)=t$. Otherwise, we can assume that it is “undefined” ($f_v(k)=-1$).

We can find array $A_v(k)$ in the same way as for $C_v(k)$ array,

($A_v(k)=0$; but if we can find $t$ which makes value $k=(\lambda_v(t)-1)/2$ an integer number and therefore $v(t)=A$, then $A_v(k)=1$). You can, by analogy with $A_v(k)$, to define array $B_v(k)$, but it is easy to show that it coincides with $C_v(k)$).

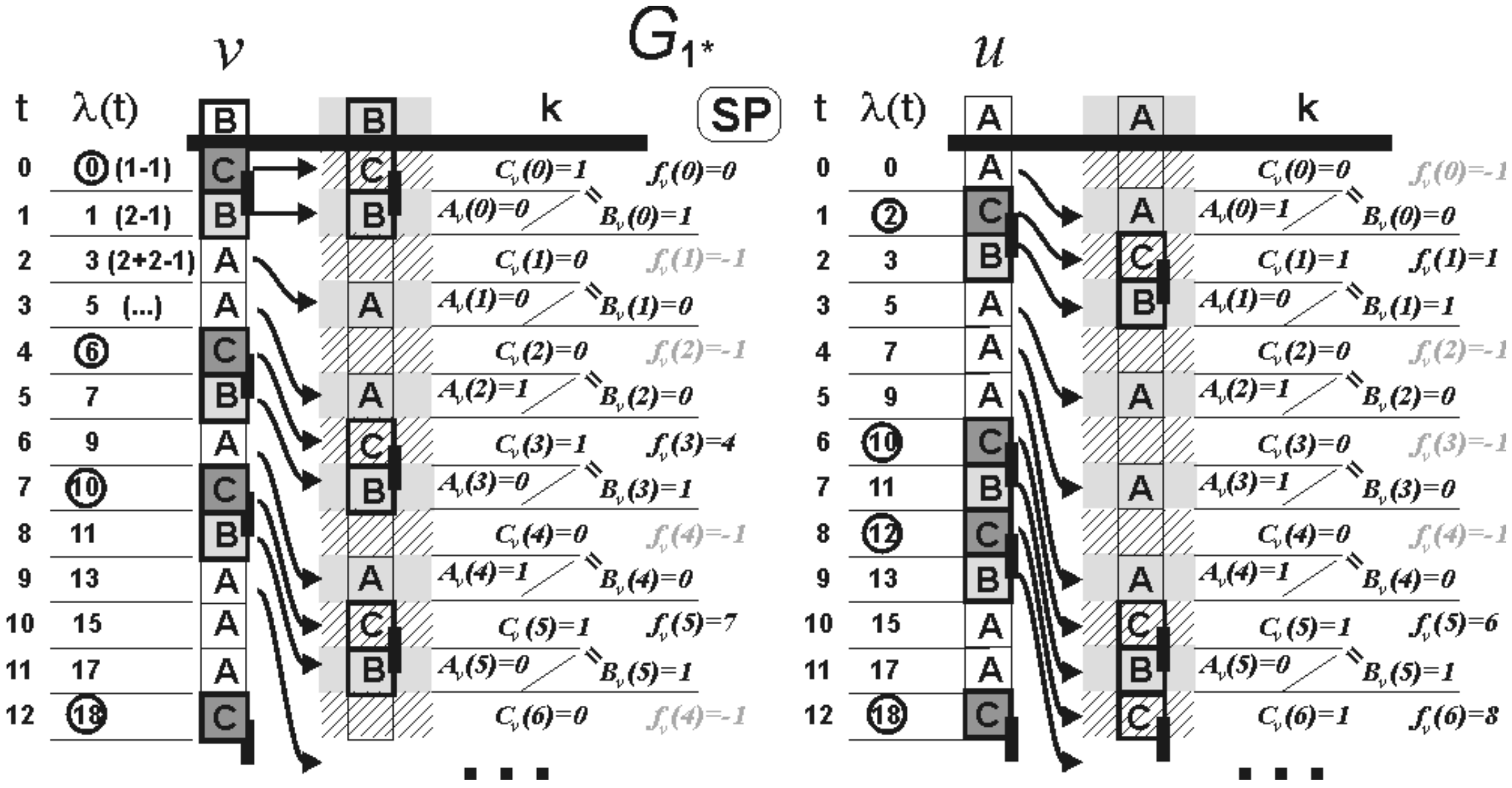

Fig. 7. Examples constructing arrays $C_v(k)$, $A_v(k)$, $B_v(k)$, $f_v(k)$. For a node $v(t)$ with the beginning: B(SP)CBAACBACBAAAC and for a node $u(t)$ with the beginning: A(SP)ACBAAACBCBAAC. From these tapes we made their secondary tapes in which the cells A are transferred "through" the empty cell. Because C and B are always in pairs λ(t) for the cells C are always obtained even (circled in figure).

Now , we can define following arrays for coloring $\overline{G}_{1*}$:

$\overline{a}_v(t), \overline{b}_v(t), \overline{c}_v(t), \overline{\lambda}_v(t), \overline{A}_v(t), \overline{C}_v(t), \overline{f}_v(t)$.

And, finally, our “impotant definitiion”…

Consider the case where value $T+\overline{T}$ can be divided by 3 with no residual, so $K=(T+\overline{T})/3$ and following

conditions are fulfilled

First three conditions are:

$$G_T = \overline{G}_{\overline{T}} \qquad [1]$$

$$\lambda = -\overline{\lambda} \qquad [2]$$

$$\overline{T} - T = \lambda \qquad [3]$$

Then next four conditions are:

$$C_v(k) + \overline{C}_v(k) = 1 \qquad [4]$$

$$A_v(k) + \overline{A}_v(k) = 1 \qquad [5]$$

$$\overline{A}_v(k) = C_v(k) \qquad [6]$$

$$\overline{C}_v(k) = A_v(k) \qquad [7]$$

for any node $v$ and $k = 0..K-1$

Additionally, from ratio [4] we can define full integral phase with modulus 2

$$F_v^{(2)}(k) = \begin{cases} f_v(k) \bmod 2 & \text{if } f_v(k) \neq -1 \\ 2 + \overline{f_v}(k) \bmod 2 & \text{otherwise} \end{cases}$$

We can also write the "eighth" condition:

$$F_v^{(2)}(0) \equiv 0 \pmod 2$$
$$F_v^{(2)}(2k\text{-}1) \equiv F_v^{(2)}(2k) \pmod 2$$
$$\textit{and if K is even}$$
$$F_v^{(2)}(K\text{-}1) \equiv F_u^{(2)}(K\text{-}1) \pmod 2$$

for all nodes $v$ and $u$ and for any $0<k<(K+1)/2$.

If all the above conditions are fulfilled, then we can make a conclusion that graphs $G_{1*}$ and $\overline{G}_{1*}$ move from SP to MP with conservation of the **Invariant of Precise Filling (IPF)**.

## 3. "X-problem of number 3" in one dimension. Definition

Consider that L nodes of graph are placed on a circle and numbered as x = 0, 1, …., L-1.

The edges of the graph are defined by two integer numbers: *n* (conventionally "left") and m (conventionally "right").

There exist only edges (x,(x-(i+1)*$b_i$(n)) mod L) and (x,(x+ (i+1)*$b_i$(m))mod L) for all *x* and *i* when funcnion $b_i$ is significant. Here, $b_i$(n) is the *i*-th bit of binary expansion of number *n*; n=$b_0$(n)+ $b_1$(n)*$2^0$ + $b_2$(n)*$2^1$…+ $b_p$(n)*$2^p$). "Mod L" shows that the circle is continuous and closed (no gaps). Let us refer to a such a construction as **Automat of Configuration 2x3** (AC23) in one dimension.

Suppose that integers *n* and *m* are odd. Then, our Automat is Weak computable. (Because it contains all undirected edges (x,(x+1)modL). It is also easy to see that all the mask containing bits $b_p$(n), $b_{p+1}$(n), $b_p$(m), $b_{p+1}$(m) = 1, that is. mask (6.7), (6.14) (6.22) - also have this property).

Suppose that there exist such *n* an *m* that Automat **at any L** and **any starting colors** $G_{1*}$, and $\overline{G}_{1*}$ always goes from SP to MP **while keeping IPF constant**. Let us call it a **correct mask**. In other case, the mask is **incorrect**.

From here we can constitute the **"X-problem of number 3" in one dimension.**

We can prove (or disprove) **that the masks (1.1). (1,3), (3,5), (3,3), (5,5) are correct**. (For example, we know that mask (1.5) is incorrect). Full list (more precisely, apparently full list) of correct masks till the values *n*, *m* = 39 is shown in Fig. 8.

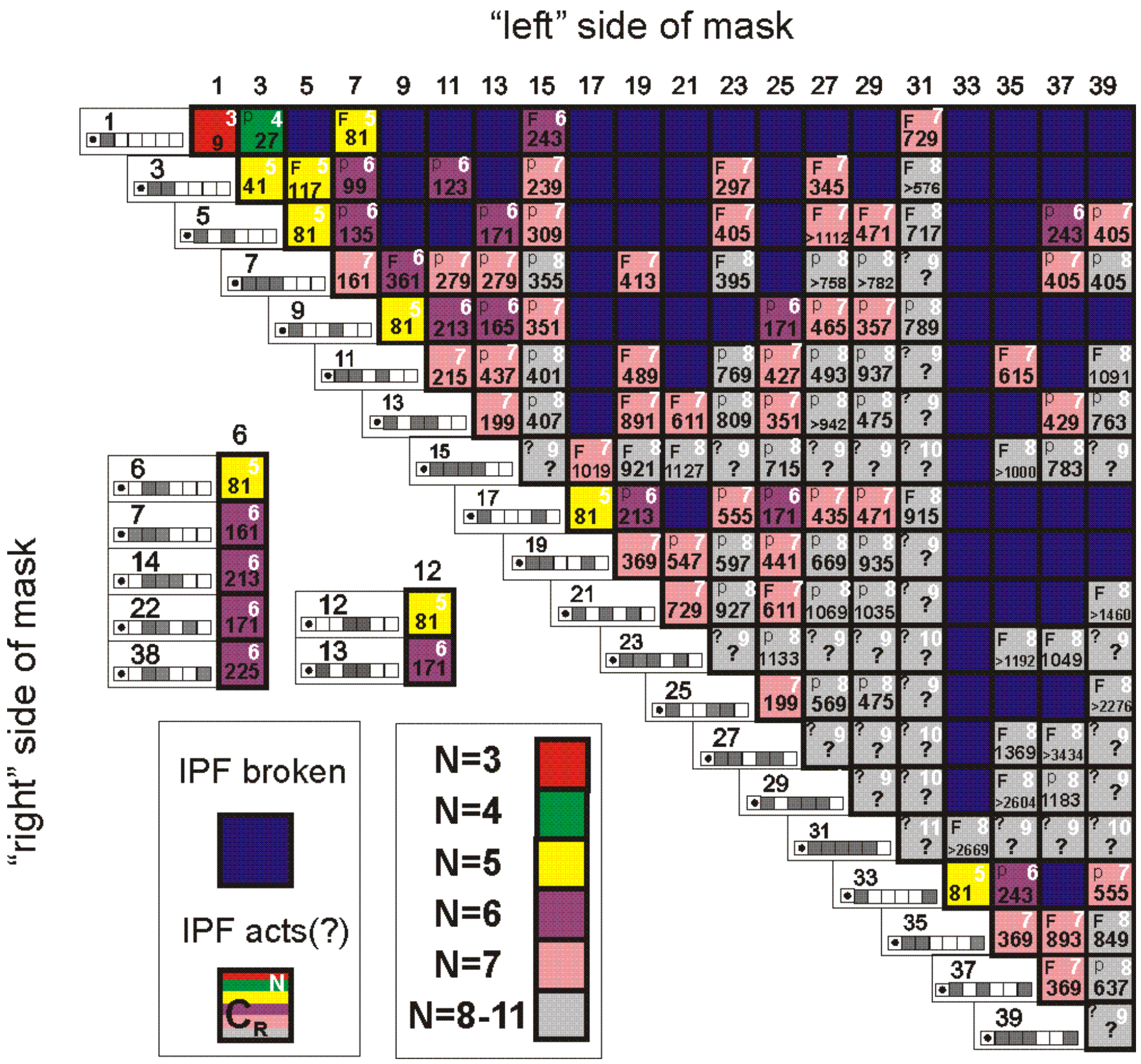

Fig. 8. Table showing which of Weak computable masks is correct(?) (shown in different colors depending on the N (total number of pixels in the mask including central point)) or incorrect (shown in dark blue). All centrally symmetric cells (the cells on the main diagonal of the table) are correct(?). In some cells, there is a designated $C_R$ value (see below).

First, we illustrate the "number 8" relation for the correct masks (see Fig. 9).

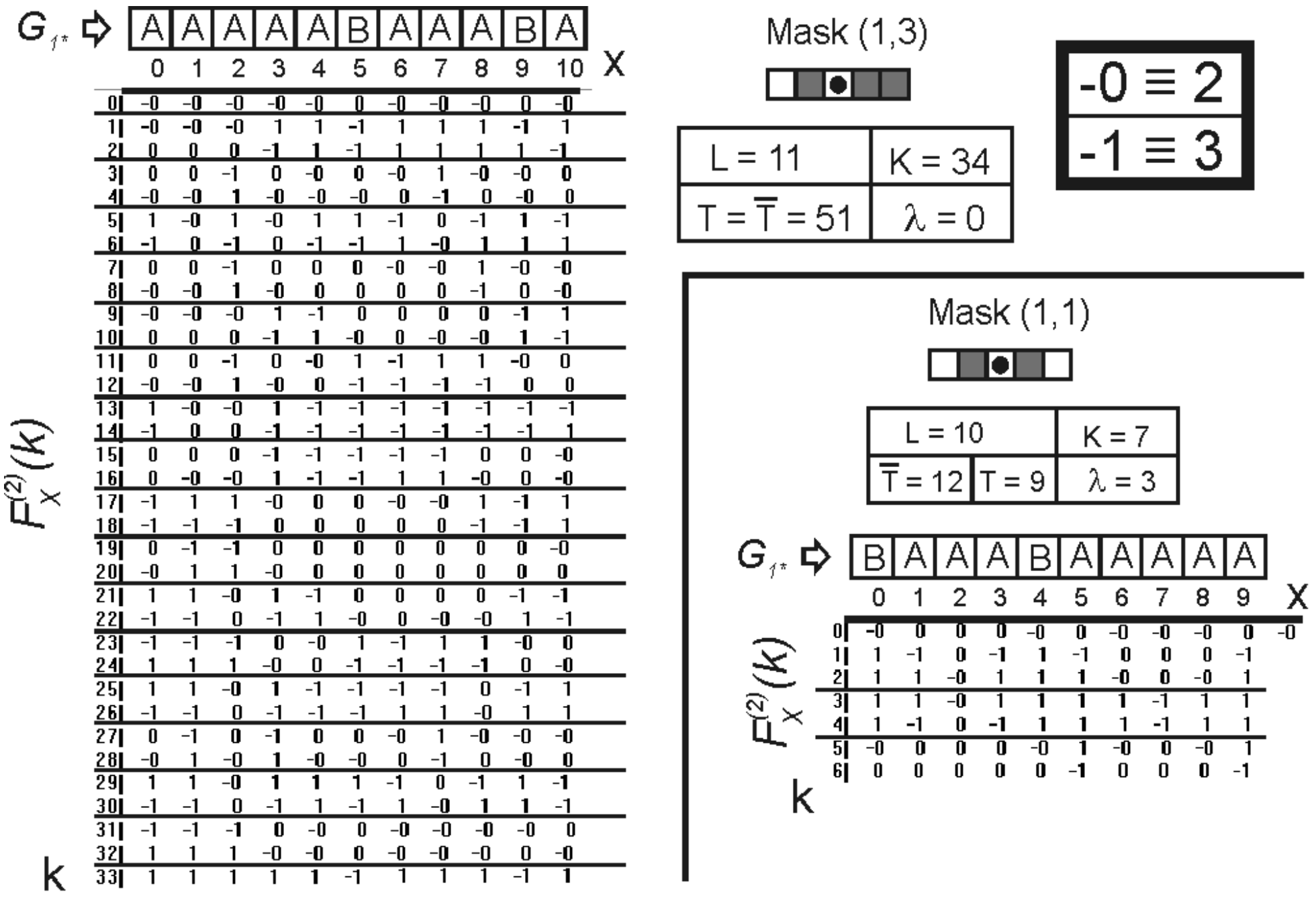

Fig. 9. Illustration of number 8 relation for correct masks (1,3) and (1,1).

It is evident that after the first row, the filling if the table is going by “pairs of lines”. In Fig. 11, the number of the last line of the mask (1.3) is odd. It is not necessary that the last line has to have the odd number. The last line number can be even.

Let us draw our attention to outgoing and very powerful observation, which, perhaps, will help us to solve the "X-problem" some day.

Let us consider a complete integrated phase with modulus 3. So, let function $F^{(3)}_x(k)$ be defined as

$$F_x^{(3)}(k) = \begin{cases} f_x(k) \bmod 3 & \text{if } f_x(k) \neq -1 \\ 3 + \overline{f}_x(k) \bmod 3 & \text{otherwise} \end{cases}$$

Then it turns out that numbers in any line (except the first one) are unambiguously determined from the previous line, more precisely, from *the numbers occupying the points (positions) directly above the corresponding points (positions) in the line of interest*. Here, the points of the mask include also a "central" point. The number of all configurations leading to the values 0, 1 and 2 *are equal* (see Fig. 12 . We refer to this algorithm as R**esolution**; then we refer to the corresponding table with the numbers 0, 1, 2 as R**esolution** T**able** (R); finally, we refer to the number of rows in this table ($C_R$) as R**esolution** C**onstant** for each correct mask.

So, having the first line and the **Resolution Table**, we can straightforwardly and without difficulty restore the entire function $F^{(3)}_x(k)$. Therefore, the same can be done for matrices $C_v$ and $A_v$. The criterion that we have reached the end (that is, we have reached MP), is precisely that $F^{(3)}_x(k) \bmod 3 = 0$ along the entire line (i.e., for all x).

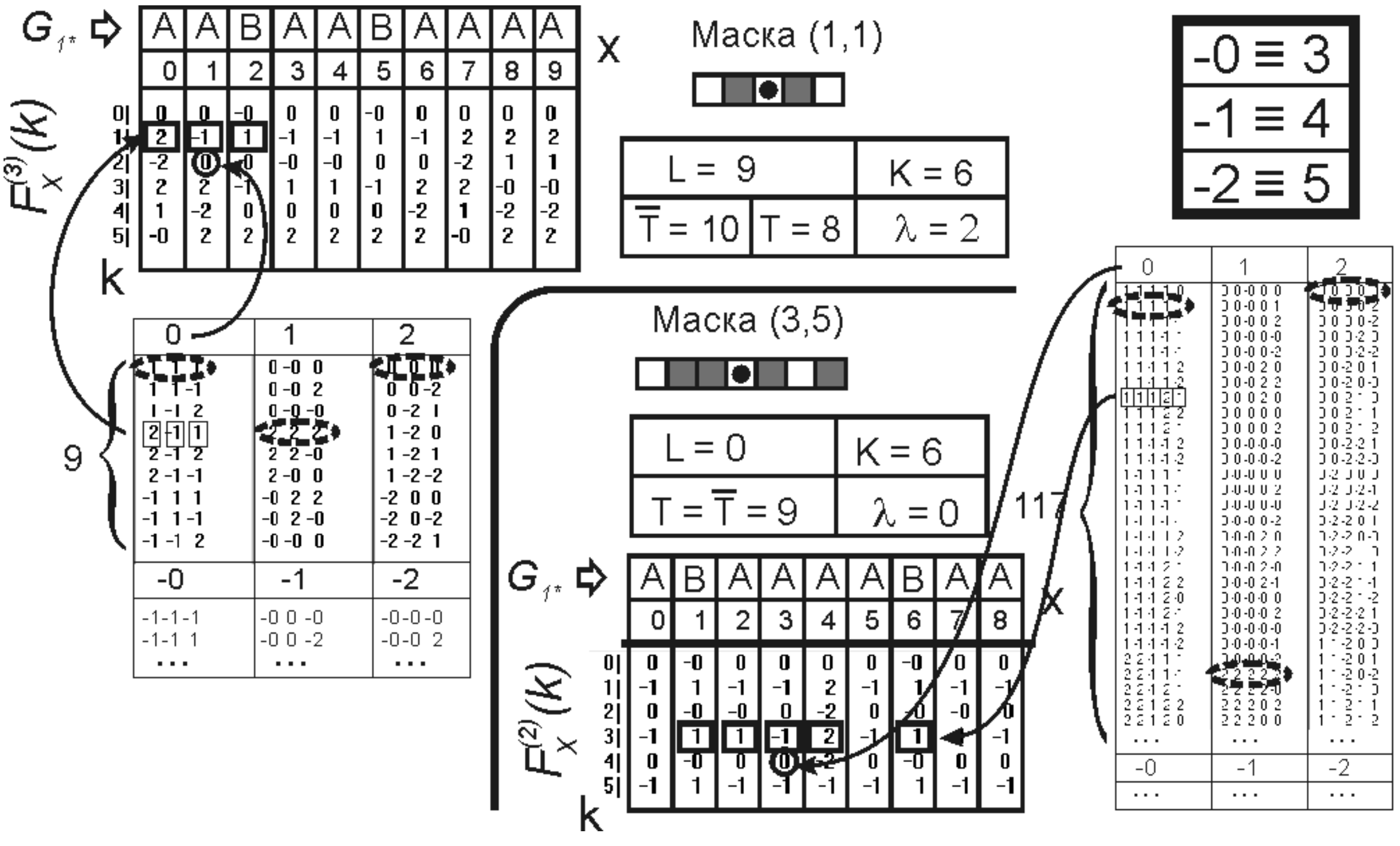

Fig. 10. Illustrations to main observations made for masks (1,1) and (3,5). This figure shows the Resolution Table for the mask (1,1) and initial lines of the Table for mask (3.5) that contains numbers 0, 1, 2 only. Tables for the numbers -0 (3), -1 (4), and -2 (5) can be obtained automatically, as "complementary" to the original.

# 4 Resolution Table for correct masks in one dimension. The first observations

1). First and foremost observations.

It is clear that the Resolution Table has a very simple symmetry. In case we have only one of the six (=0, =1, =2, =3(-0), =4(-1), =5(-2)) "sub-tables", we can determine the five remaining sub-tables too. To do this, we make a substitution as shown in Fig. 11. (In Fig. 11, the "base number" for each of the sub-tables is shown in gray). There is a variant for each sub-table that contains the "base number" in all the cells. Such variants are indicated by a dashed-line oval in Fig 10).

| =0 | =1 | =2 | =$\bar{0}$ | =$\bar{1}$ | =$\bar{2}$ |
|---|---|---|---|---|---|
| 1 | 2 | 0 | $\bar{1}$ | $\bar{2}$ | $\bar{0}$ |
| $\bar{1}$ | $\bar{0}$ | $\bar{2}$ | 1 | 0 | 2 |
| 2 | 0 | 1 | $\bar{2}$ | $\bar{0}$ | $\bar{1}$ |
| $\bar{2}$ | $\bar{1}$ | $\bar{0}$ | 2 | 1 | 0 |
| 0 | 1 | 2 | $\bar{0}$ | $\bar{1}$ | $\bar{2}$ |
| $\bar{0}$ | $\bar{2}$ | $\bar{1}$ | 0 | 2 | 1 |

Fig. 11.

So let's take a "subtable" to output the numbers (=0), and we just call *it* a Resolution Table.

So R – is the set of $C_R$ numbers $\{a_0, a_1 \ldots a_{N-1}\}$, where $a_i \in \{0,1,2,3,4,5\}$ (For numbers 3, 4, 5 will still use the notation, or $\bar{0}$, $\bar{1}$, $\bar{2}$ or 0, -1, -2, where both is comfortable).

We have an "isolated position" in our mask: a central point. Let's put her column number 0 in the Resolution Table. And to have all the other numbers of colomns we just list the points of the mask from left to right. (So for the mask (3.5) we obtained: "column 0" in the table – the central point (offset 0). For "column 1" – point with offset 2. For "column 2" – point with offset -1. Further, for the "3 column" (jump over 0) – the point with offset 1. And for the last column – the point with offset 3).

There is one more symmetry. If the mask reflect by central point, and keeping the same correspondence between the column and the points of the mask: that is the column witch offset x now assign a point with offset -x ... then we come to the identity same Table. Probably, this symmetry is trivial.

2). The masks is devited into three classes on the subject: what numbers, in general, are in their Resolution Tables.

1. *Small* class (contains only the numbers 1, 2, 4(-1)). "Nothing" in the top left corner of Fig. 8.

2. *Middle* class (contains the numbers 1, 2, 0, 4(-1)). Letter «p» ("partial") in the top left corner of Fig. 8.

3. *Full* Class (contains all the numbers 0, 1, 2, 3(-0), 4(-1), 5(-2)). Letter «F» in the top left corner of Fig. 8.

Small class includes (apparently?) all centrally symmetric mask, at least for which n = m is less then 33.

With the increase of n and m mask more and more pass away in a full class. But there are a lot of exceptions.

We denote the total number being in RT values 0, 1, 2, 3, 4, 5 *except zero column* as $s^{+0}$, $s^{+1}$, $s^{+2}$, $s^{-0}$, $s^{-1}$, $s^{-2}$.

We denote the total number being in the RT values 0, 1, 2, 3, 4, 5 in the zero column as $s_0^{+0}$, $s_0^{+1}$, $s_0^{+2}$, $s_0^{-0}$, $s_0^{-1}$, $s_0^{-2}$.

All of these values for all masks (n, m) with odd n, m <19 shown in Figure 12.

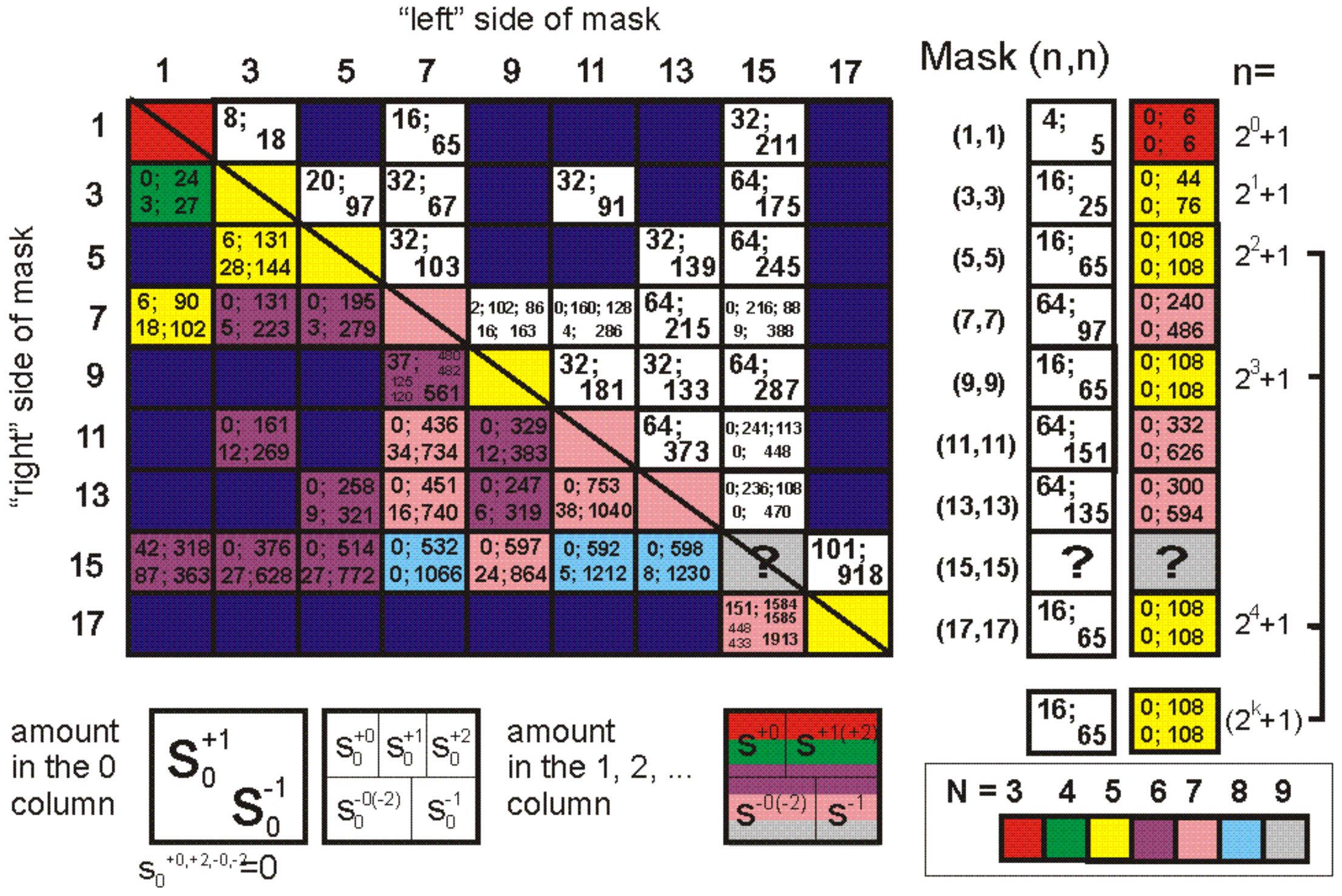

Fig. 12. Summary ("s") of the numbers 0, 1, 2, 3, 4, 5 in RT. (N-total number of points in the mask, including the center).

We can say that the correct masks are divided at the *completely* correct (ie, those for which $s_0^{+0.+2,-0,-2}$ =0) and *not completely* (otherwise).

In turn, a completely correct masks are divided into completely correct of the 1st kind (ie those for which $s_0^{+1} = 2^{N-1}$ and sub-table with zero column is equal to a imple enumeration of all possible combinations of +1 and -1, and of the 2st kind (otherwise).

Completely correct mask of the 1st kind – all symmetric masks (at least for n <19), and a mask (1,3), (1,7), (3,7), (5,7) (3.11) ...

Completely correct the mask of the 2nd kind: masks (3.5), (15.17) ...

We denote the RT of the mask (n, m) as [n, m], simply by changing the brackets round to square.

Once we find the identity [5,5] = [9,9] = [17,17] = ... = $[2^k +1, 2^k +1]$ where k > 1.

4). Building Resolution Tables for masks of type $(1,2^k-1)$.

First, we must numbering the columns. Number them in accordance with tye paragraph 1 of this chapter. (Central point – column number 0, etc. …)

Here is how we sort our rows. Assume that we have a row (**-1**, 2, 0, -2, 1). We transfer the column № 0 of the beginning to the end. Get string (2, 0, -2, 1, **-1**). And consider this string as a record number in a sextuple number system, in which the bits are counted from right to left. (2, -0, -2, 1, **-1**) = 4 +1*6 +5*$6^2$ +3*$63^3$ +2*$6^4$ = 4 +6 +180 +6048 +2592 = 8830. And now we place the rows in ascending of these numbers.

Now Resolution Table for masks of type $(1, 2^k-1)$ can be easily constructed by induction of k. See Fig. 13.

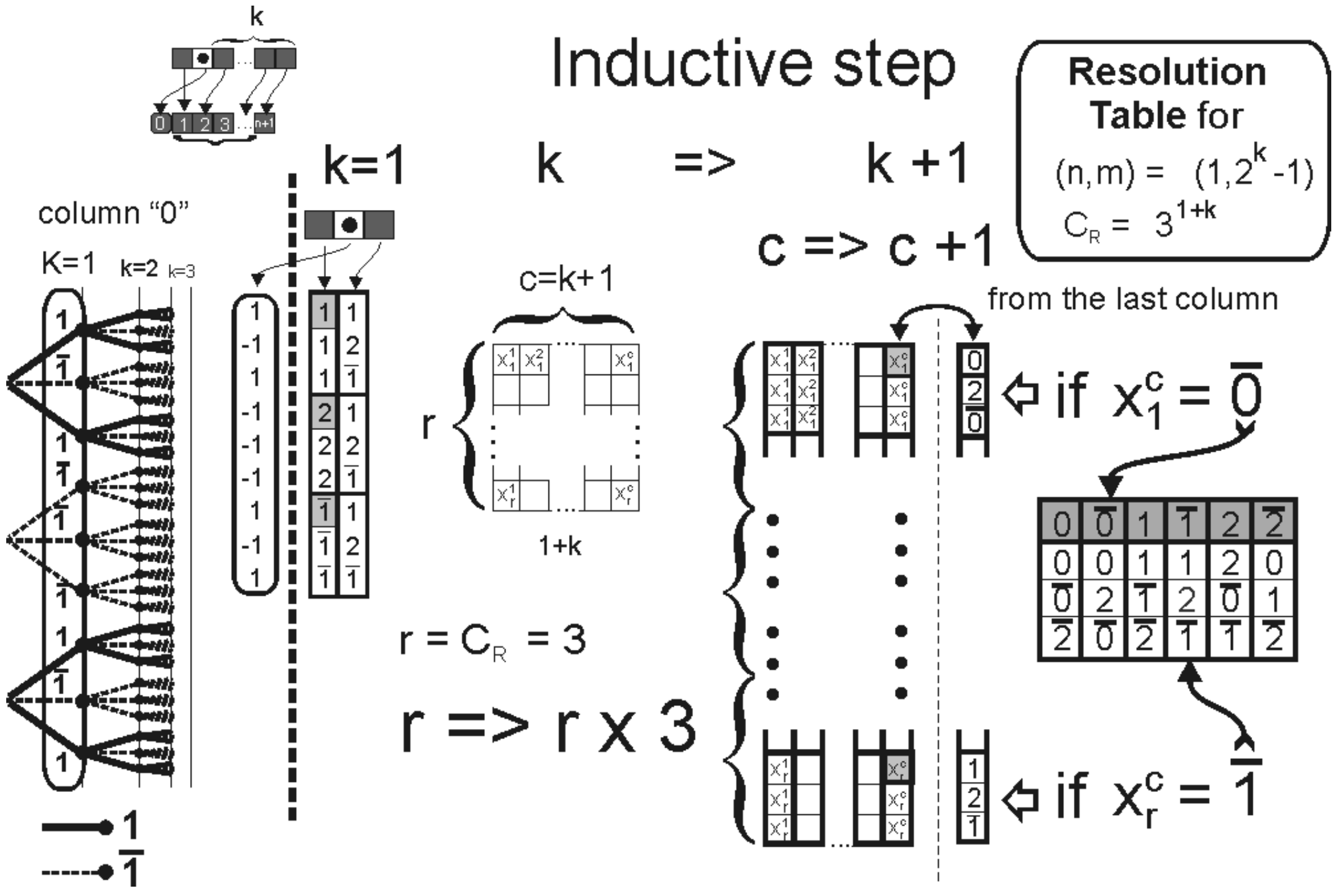

Fig. 13. Building a Resolution Tables for mask $(1,2^k-1)$ by induction.

Zero column is obtained by simple linear fractal of the numbers 1 and -1 (see Fig. 13, left).

The value of Resolution Table for k = 1 is shown.

Now the induction step for the other columns.

For him, the number of lines is expected to increase by 3 times ($C_R$ increases three times), and the number of columns increase by one.

We operate as shown in the figure. The table "triples" trivial way, and the new column is obtained from the latter in accordance with the table below. We can show that this will always turn out to completely correct mask of the 1st kind.

Also it can be said that we obtain all Resolution Tables for *solid* masks - masks form $(2^k-1, 2^p-1)$ where k, p> 0. Because RT mask $(1,2^{k+p}-1)$ *includes* the entire series with constant k+p. But more on that in the next paragraph...

4). Our final observation concerning the *compatibility* of Resolution Tables for various correct masks with the same number of columns.

So, we have to take the mask with the same N (N – recall, the total number of pixels in the mask, including the central), and check each rows of the association of Resolution Tables for the question: does its converted rows appear in some different sub-tables which we have. If there is – then masks are not compatible.

Our observation: if we take the numbering of the columns in accordance with paragraph 1 of this chapter then all correct mask with the same N *will remain compatible*.

That is, for each N there is its own "integral" Table $R^{(N)}$, with a corresponding number of lines in it. Observation has been tested for each N <7.

To begin with N = 4. With such N there are only two right masks (1,2) and (2,1). (See Fig. 14). Their $C_R$=27. $C_R$=37 is their association. (Resolution Table intersect at 17-strings). Remember, to delete the zero column is STRICTLY NECESSARY! Otherwise, even in the case of N = 4 there will be numerous violations!

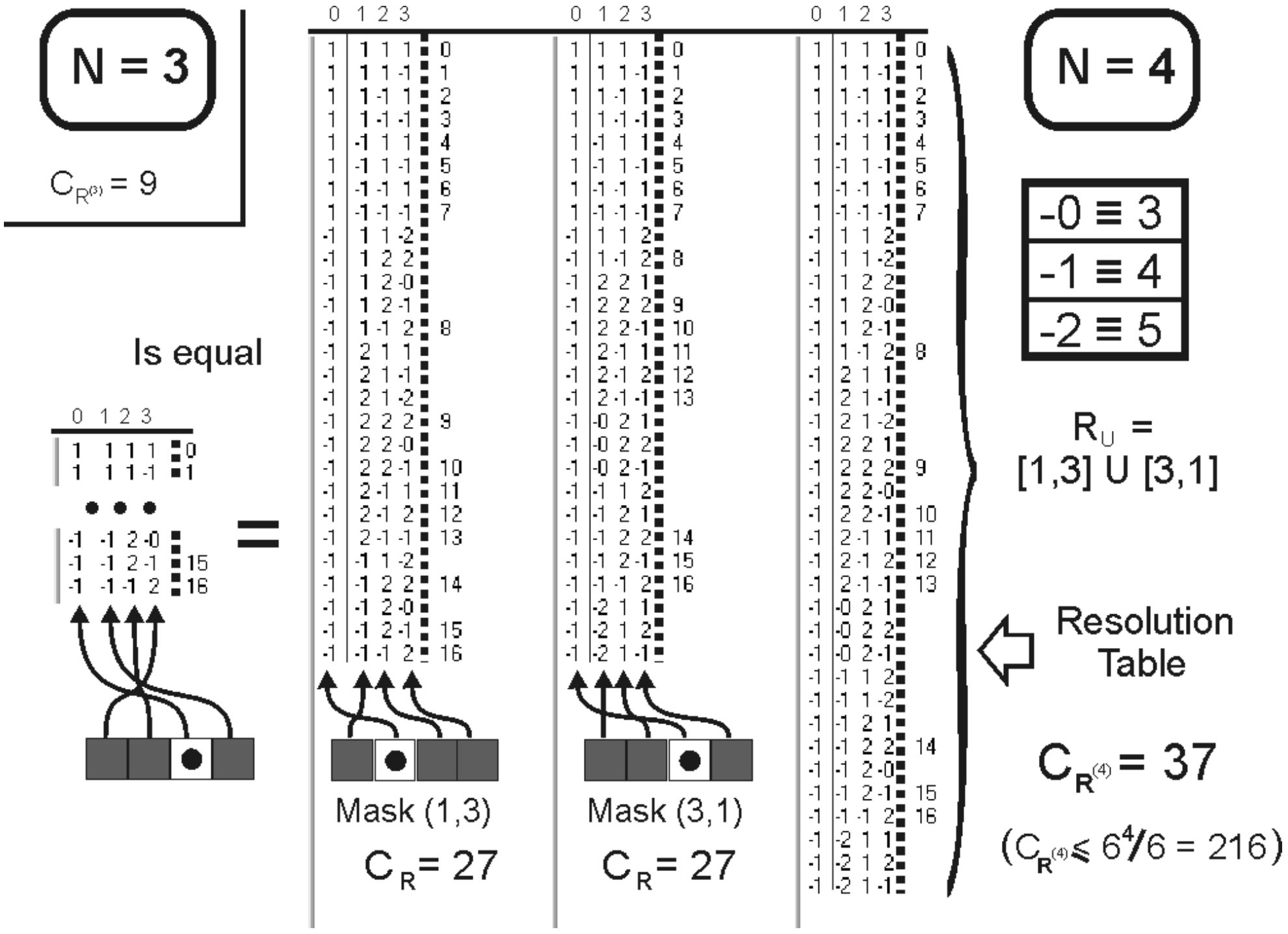

Fig. 14. The construction of the integrated RT for N = 4.

(Note again: if we take the far left column numbering in Fig. 14, according the "trivial" symmetry in paragraph 1 (the column with the shift x for mask (1,3) corresponds to a column with a shift -x for mask (3,1)), then we would come to RT identical to the first table. Our association is not trivial).

Consider the case N = 5. See Fig. 15. Apparently all of the correct and "obvious Weak computable masks" depicted on it.

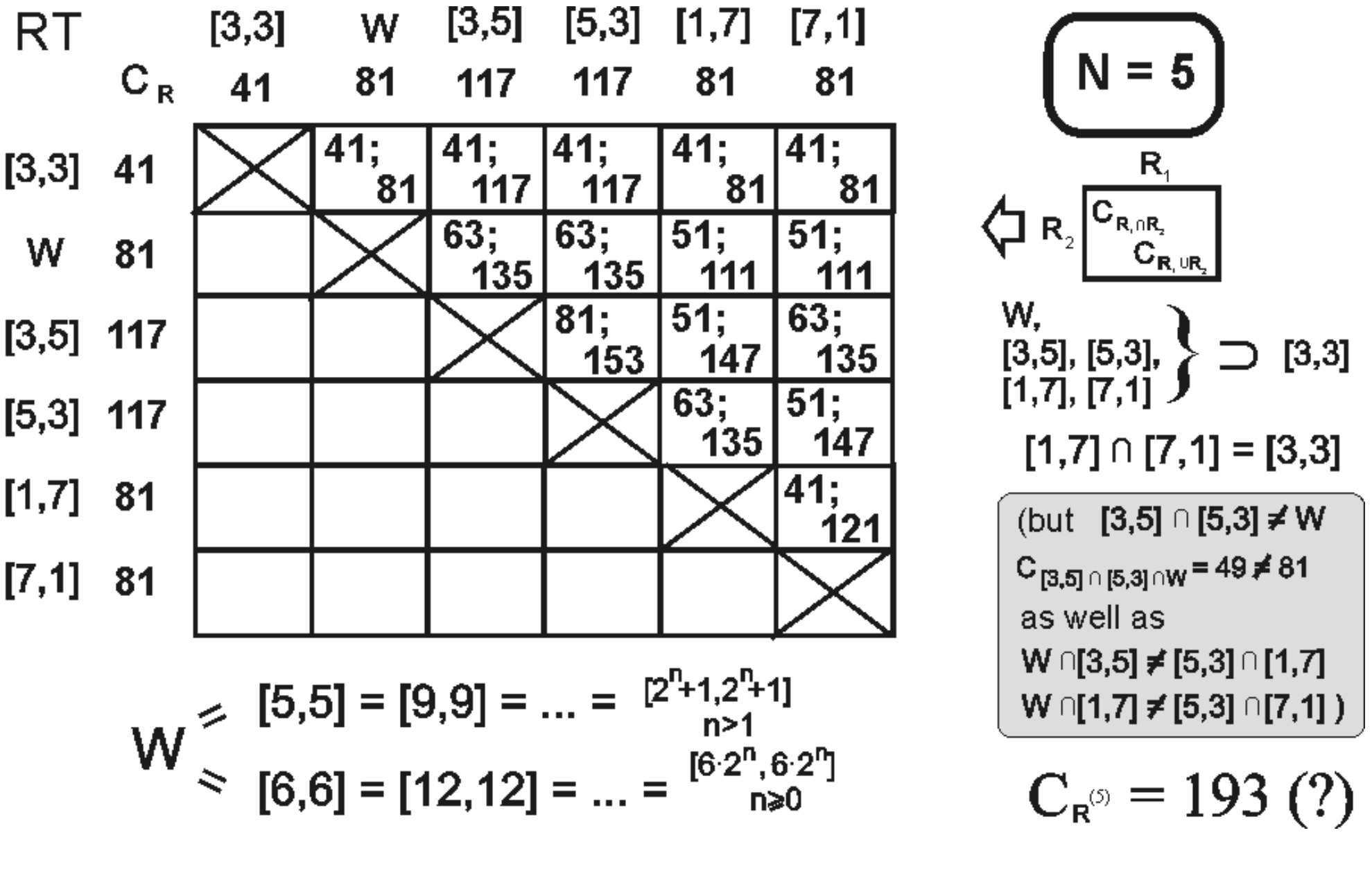

Fig. 15. The construction of the integrated RT for N = 4. In the cells of the table is: in the upper left corner – CR for the intersection of RT masks, and at the bottom right corner – to combine.

They consists of two infinite series of equal RT masks W = [5,5] = [9,9] = [17,17] ... = [6,6] = [12,12] = [24,24] ... plus five remaining correct [3,3], [3,5], [1,7], [7,1]. (All other possibilities: masks (3.9), (3, 17) ... and (5.9), (5.17) ... - uncorrect). However, we have put close to integral number 193 the question mark.

The case with N = 6 is much more diverse. See Figure 16.

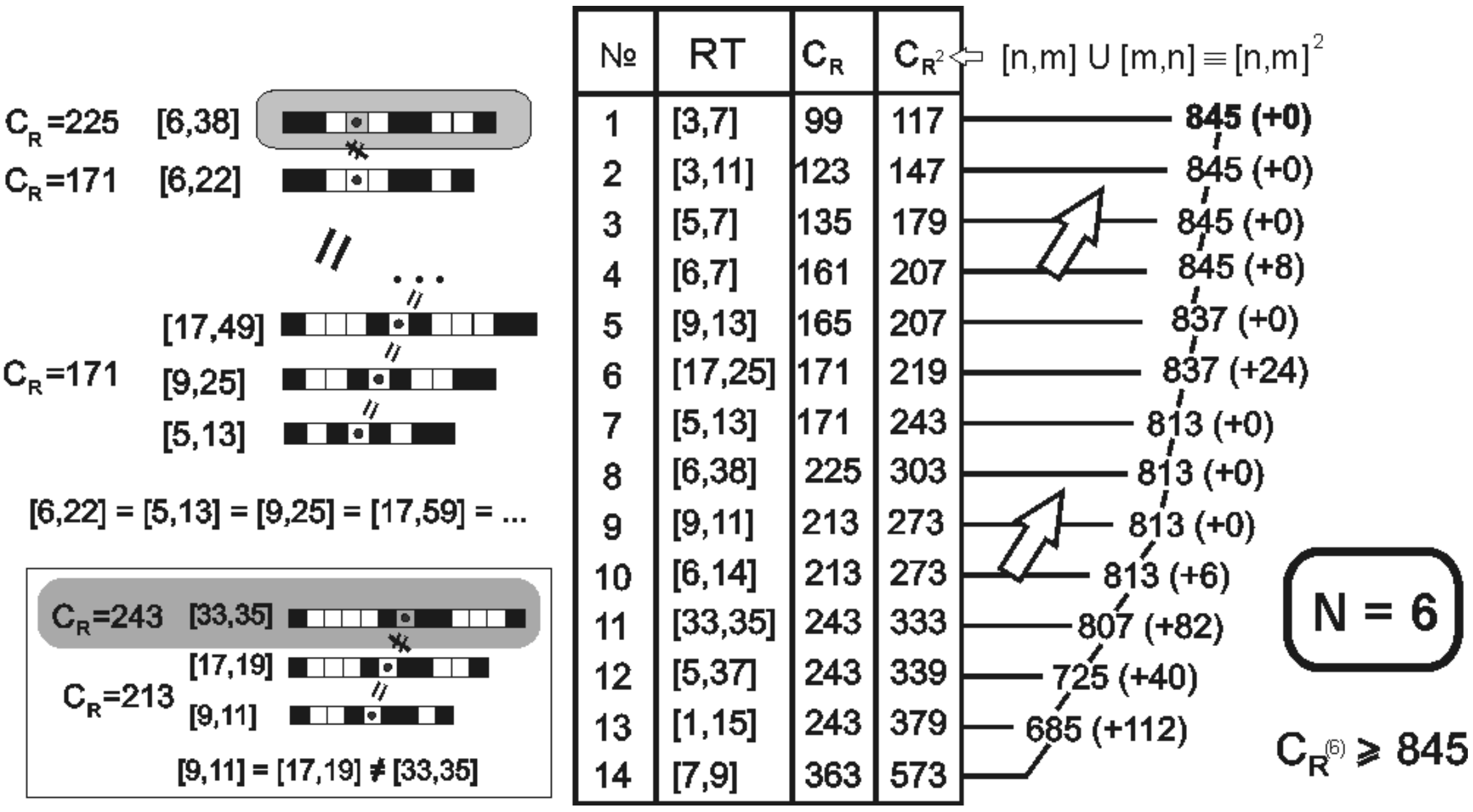

Fig. 16. Three identities for masks with N = 6. Right - the steps of calculating the integral $C_R$ .

It has a (seemingly endless) line of equal RT: [5,13] = [9,25] = [17,49] ... With this line coincides RT [6,22]. But, unlike the case with N = 5, there is not second line. (See picture 16). Still, there is one "stand-alone" identity [9,11] = [17,19], which also does not have a continue.

The right side shows the steps of providing an integrated table $R^{(6)}$. We have used the correct mask with N = 6 and odd n, m <40 and have added three correct mask with n = 6. Then sort them in ascending $C_R$. After that, every RT [n, m] was combined with it centrally symmetric RT [m, n]. And then started to fold, starting with the largest $C_R$. The results in Fig. 16. The resulting final value (845) is hardly much less then accurate.

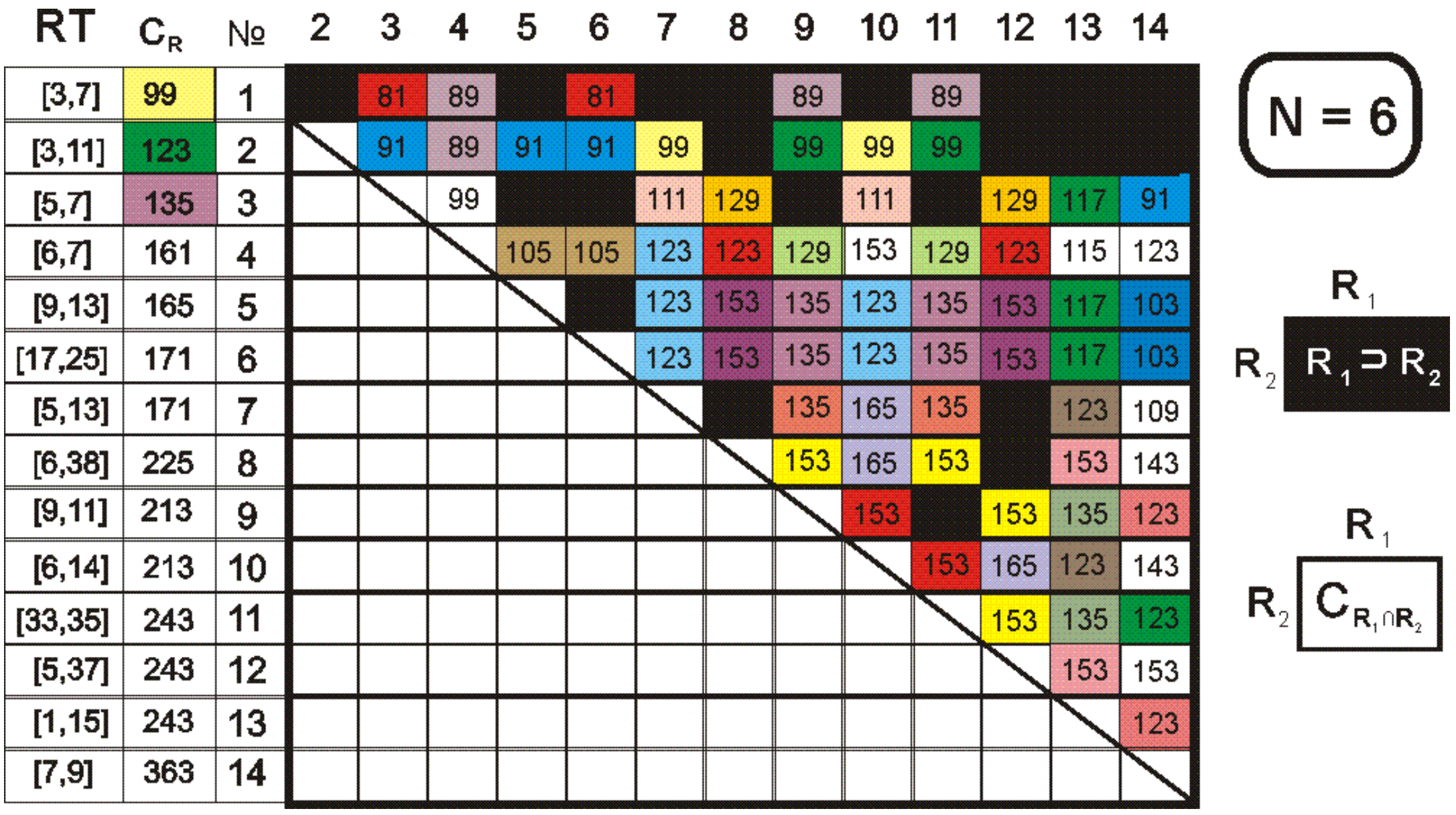

Fig. 17. "Table of coincidence" for Resolution Table for masks with N = 6.

Figure 17 shows the "table of coincidence" for the RT of our masks. If the cell is black, it means that one mask includes another (upper includes the left). Otherwise, in cells are $C_R$ (number of rows) in the intersection table. If these values are the same and the cells are painted in the same color (but not white!), it means that not only the number of rows, but also the RT itself is the same. Examples: [3,11]=[33.35]∩[7.9]; [6.38]∩[9.11]=[33.35]∩[5.37]; [6.7]∩[9.13 ] = [17.25]∩[6.14] ... and so on.

It can be seen that the algebra of Resolution Tables is very intricately.

## 5 Conclusion

There is no doubt that similar "problem of number 3"exists *in all dimensions* [2].

Using a personal computer (PC), we tested the following amazing statement: each and every **mask in one, two, and three dimensions**, which includes all the points adjacent to the "central" point (we refer to such mask as simple) and demonstrates the property of central symmetry (that is, is symmetric with respect to the central point) **is correct**.

After several months of PC operation, we did not find a single exception! (We used the "light" test, that is, we checked only the first three of the nine statements. We found that building *F* arrays in two and three dimensions has been rather challenging. So, it is impossible to guarantee that this statement is true. Some other interesting aspects of the "X-number of problems 3" in two dimensions can be found in paper [3].

Readers can download illustrative program from kornju.hop.ru .

**Acknowledgments**

**We thanks Dr Eugene Moskovets for reading and translating this manuscript.**